\documentclass[pdflatex,sn-chicago]{sn-jnl}

\usepackage{graphicx}%
\usepackage{multirow}%
\usepackage{amsmath,amssymb,amsfonts}%
\usepackage{amsthm}%
\usepackage{mathrsfs}%
\usepackage[title]{appendix}%
\usepackage{xcolor}%
\usepackage{textcomp}%
\usepackage{manyfoot}%
\usepackage{booktabs}%
\usepackage{algorithm}%
\usepackage{algorithmicx}%
\usepackage{algpseudocode}%
\usepackage{listings}%
\usepackage{tabularx} %
\usepackage{comment} %
\usepackage{pdflscape} %
\usepackage{hyperref}

\theoremstyle{thmstyleone}%
\theoremstyle{thmstyletwo}%

\theoremstyle{thmstylethree}%

\begin{document}

\title[Article Title]{Does Scientific Writing Converge to U.S. English?\\
Evidence from Generative AI-Assisted Publications}


\author*[1]{\fnm{Dragan} \sur{Filimonovic}}\email{dragan.filimonovic@unibas.ch}

\author[1]{\fnm{Christian} \sur{Rutzer}}

\author[2]{\fnm{Jeffrey T.} \sur{Macher}}

\author[1]{\fnm{Rolf} \sur{Weder}}

\affil*[1]{\orgdiv{Faculty of Business and Economics}, \orgname{University of Basel}, \orgaddress{\street{Peter
Merian-Weg 6}, \city{Basel}, \postcode{4002}, \country{Switzerland}}}

\affil[2]{\orgdiv{Robert E. McDonough School of Business}, \orgname{Georgetown University}, \orgaddress{\street{37th
and O Streets NW}, \city{Washington}, \postcode{DC 20057}, \country{USA}}}


\abstract{
A growing literature documents that generative artificial intelligence (GenAI) is changing scientific writing, yet most studies focus on absolute changes in vocabulary or readability. An important question remains unanswered: Does GenAI use lead to systematic convergence, or a narrowing of stylistic gaps relative to the dominant form of scientific English? Unlike absolute changes, convergence signals whether language-related publication barriers are declining and suggests broader implications for participation and competition in global science. This study directly addresses this question using 5.65 million English-language scientific articles published from 2021 to 2024 and indexed in Scopus. We measure linguistic similarity to a U.S. benchmark corpus using SciBERT text embeddings, and estimate dynamic changes using an event-study difference-in-differences design with repeated cross-sections centered on the late-2022 release of ChatGPT. We find that GenAI-assisted publications from non-English-speaking countries exhibit statistically significant and increasing convergence toward U.S.\ scientific English, relative to non-GenAI-assisted publications from these countries. This effect is strongest for domestic author teams from countries more linguistically distant from English and for articles published in lower-impact journals---precisely the contexts where language barriers have historically been most consequential. The results suggest that GenAI tools are reducing language-related barriers in scientific publications. Whether this represents genuine inclusion or a deepening dependence on a single linguistic standard remains an open question.}

\keywords{generative AI, linguistic convergence, scientific writing, SciBERT, bibliometrics, difference-in-differences}
\pacs[JEL Classification]{O33, D83, C23, C55, I23}

\maketitle

\section{Introduction}

The dominance of English in the global scientific community has long conferred structural advantages on native English-speaking researchers and created persistent obstacles for non-native English-speaking researchers \citep{Montgomery2013,Amano2016,nature2023scientific}. Non-Anglophone researchers often face greater writing burdens and peer-review bias risks associated with language proficiency, with well-documented publication success and career progression effects \citep{Clavero2010,Hanauer2019,Politzer2020,Castaneda2020,Amano2023}. These disadvantages have historically required substantial investments, including time spent in English-speaking research environments and payments made to professional language editing services \citep{Stephan2016,Balan2021}.

The late-2022 release of ChatGPT, however, introduced a widely-accessible and low-cost alternative for text editing and language assistance---offering a potential means of reducing these longstanding linguistic barriers \citep{Bubeck2023,van2023}. A growing body of bibliometric and empirical research has subsequently documented the rapid uptake of generative artificial intelligence (GenAI)-assisted scientific writing and the associated shifts in vocabulary use and textual complexity \citep{Geng2024,Gray2024,Liang2024,Uribe2024,dehaan2025gpt,feyzollahi2025adoption,kobak2025llm}. Non-native English-speaking and less-established researchers are the most active adopters of GenAI writing tools \citep{Lin2025,Liu2025}, and productivity gains are substantially larger for researchers at non-Anglophone institutions \citep{Kusumegi2025}. Readability improvements among non-native English-speaking researchers are also documented, with particularly pronounced gains in countries more linguistically distant from English \citep{Prakash2025}. 

These contributions, however, share a common limitation: in particular, they measure absolute changes in writing features, such as vocabulary shifts or readability scores, rather than convergence toward a specific linguistic benchmark. This distinction is not merely methodological but conceptually meaningful: viz., absolute improvement indicates that writing is changing; convergence indicates where writing is moving and, by implication, which and whose norms are being adopted. For example, a publication may become more readable without becoming more similar to U.S. English, or it may converge toward a dominant register yet show no aggregate readability gain. Disentangling these mechanisms requires a comparative design in which non-U.S. scientific publications are evaluated against a well-defined (e.g., U.S., English, scientific) benchmark. No existing study of which we are aware provides such an analysis at scale across published journal articles and multiple scientific fields. The extant research instead focuses predominantly on preprint repositories \citep{Kusumegi2025} and on narrowly-defined scientific fields \citep{Lin2025,Liu2025,Prakash2025}, leaving unanswered whether patterns generalize across the peer-reviewed scientific record.

To address this research gap, this paper provides large-scale evidence of linguistic convergence across multiple scientific fields. We track the stylistic evolution of 5.6 million English-language publications by authors from non-English-speaking countries, published from 2021 to 2024 and indexed in Scopus. For each publication, we measure linguistic similarity to a benchmark corpus of U.S.\ scientific writing using SciBERT\footnote{SciBERT is a transformer architecture based on BERT \citep{devlin2019bert} and pretrained on large-scale scientific corpora that effectively captures the semantic and stylistic properties of academic writing.} text embeddings \citep{beltagy2019scibert}. We then estimate dynamic changes using an event-study difference-in-differences design centered on the late-2022 release of ChatGPT. The results indicate a pronounced post-2022 convergence of GenAI-assisted publications toward U.S. scientific writing, relative to non-GenAI-assisted publications. The association is strongest where language barriers are greatest: namely, for domestic author teams from countries more linguistically distant from English and for publications in lower-impact journals.

This paper makes three contributions. First, we provide cross-scientific field and cross-country evidence that GenAI use is associated with measurable convergence in the linguistic presentation of published science---importantly going beyond the absolute changes documented in prior research. Against the broader backdrop of research on linguistic barriers and biases in global science \citep{Hanauer2019,Politzer2020,Castaneda2020,Clavero2010, Amano2023}, our results provide concrete evidence of a technology-driven counter-trend in demonstrating that GenAI use is actively narrowing geographic and linguistic disparities in the formal scientific record. If sustained, such a reduction in language-related entry barriers could broaden participation in international journals and diversify the pool of ideas on which future research builds. This consideration is made more pressing by recent evidence of declining scientific novelty and disruption \citep{Bloom2020, Park2023}. We note, however, that these potential gains must be weighed against the risk that convergence toward a single dominant register may entail homogenization at the cost of linguistic diversity. We consider in detail these system-level tensions in our discussion.

Second, we add to research that examines how digital technology adoption influences various academic outcomes: e.g., internet access has been shown to increase research output by lowering discovery costs and facilitating collaboration \citep{Barjak2006, Pachi2012}; higher country-level internet penetration is positively correlated with national publication volumes \citep{Xu2021}. More directly relevant to our study, prior research examines the role of pre-ChatGPT AI tools in enhancing writing quality and efficiency for non-native English speakers: e.g., \citet{Ghufron2018} and \citet{Fitria2021} find that using the AI-powered writing assistant Grammarly substantially improves students' English writing performance; \citet{Mundt2016} show that machine translation tools such as Google Translate enhances academic exchange and diversifies researcher communities. We build on these research efforts by examining how the latest generation of GenAI tools reshapes linguistic patterns in the published scientific record at a much larger scale.

Third, we complement the expanding literature on the broader research outcomes of GenAI adoption. Beyond the readability enhancements highlighted in the aforementioned studies, parallel research demonstrates how these tools reshape knowledge production across other demographic and disciplinary boundaries: e.g., \citet{Kusumegi2025} analyze three major preprint repositories to show that LLM usage substantially accelerates manuscript output and fundamentally alters traditional quality signals---noting that the historical correlation between language complexity and peer-review outcomes reverses for LLM-assisted papers; \citet{Filimonovic2025} find that GenAI adoption in the social sciences significantly boosts overall research productivity and modestly improves publication quality---particularly for early-career scholars. \citet{Tang2025} caution, however, that these benefits are not evenly distributed, and find using SSRN preprint data that male researchers’ productivity gains from GenAI currently exceed those of female researchers, thereby widening existing research gender gaps. By providing large-scale evidence of linguistic convergence toward U.S. English in scientific writing, our study introduces a geographic and linguistic mechanism through which GenAI may reshape global science participation---thereby complementing existing efforts on productivity, quality, and inequality in research outcomes.

The rest of the paper is structured as follows. Section 2 presents the data and methods. Section 3 reports the baseline estimates of the association between GenAI use and linguistic convergence toward U.S. scientific writing. Section 4 assesses empirical robustness under alternative keyword thresholds, stricter GenAI identification rules, and different U.S. benchmark corpus definitions. Section 5 discusses the findings and their implications, and Section 6 concludes.

\section{Data and Methods}\label{sec:methods}

\subsection{Data}

Our starting dataset is all scientific publications written in English and indexed in the Scopus database from 2021 to 2024 \citep{singh2021journal,elsevier2023}. For each publication, we obtain information on the publication year, Scopus pre-defined scientific field, journal impact factor, and author country (based upon institution affiliation). For authors with more than one affiliation, we use the first listed as it typically corresponds to the main institution. We drop observations for publications where all author affiliations are to English-speaking countries.\footnote{We retain observations for publications where at least one author affiliation is to a non-English-speaking country, as it allows for mixed-language author team analysis.} We focus on 5.65 million unique publications that include at least one author affiliation to a non-English-speaking country. We add observations for publications that classify into multiple countries (based upon author affiliation) and for publications classified into multiple scientific fields---yielding one observation per unique publication, country and field triad. Our unit of analysis is therefore the publication $\times$ country $\times$ field cell, comprising a sample of 14,315,948 observations.

To improve interpretability in cross-scientific field comparisons, we aggregate the detailed Scopus journal-based scientific fields into four broad groups: Life Sciences, Physical Sciences, Engineering and Technology, and Social Sciences. Multidisciplinary and Arts and Humanities journals are excluded due to their small sample size and ambiguous classification. This dimensionality reduction simplifies cross-scientific field analysis while retaining the main structural differences among these scientific domains. Table~\ref{tab:fields} provides the mapping from Scopus scientific fields to these aggregated groups. Each publication inherits the field assignments of its journal. Journals listed in multiple fields contribute to each corresponding scientific field.

\begin{table}[htbp]
\centering
\caption{Scopus Scientific Fields to Aggregated Group}
\label{tab:fields}
\begin{tabularx}{\textwidth}{lX}
\toprule
\textbf{Aggregated Group} & \textbf{Scopus Scientific Fields} \\
\midrule
Life Sciences & Biology, Medicine, Public Health, Immunology, Neuroscience, Pharmacology, Nursing, Dentistry, Veterinary Science, Agriculture. \\
Physical Sciences & Physics, Chemistry, Materials Science, Mathematics, Environmental Science, Earth Sciences. \\
Engineering and Technology & Engineering, Chemical Engineering, Computer Science, Energy Studies. \\
Social Sciences & Economics, Business, Psychology, Sociology, Decision Sciences. \\
\bottomrule
\end{tabularx}
\footnotetext{Note: This table maps Scopus scientific field classifications into aggregated groups.}
\end{table}

\subsection{Identifying GenAI-Assisted Papers}

We identify GenAI use by detecting lexical pattern shifts in publication titles and abstracts before and after the launch of ChatGPT. Our approach builds on recent evidence that text produced with the assistance of large language models (LLMs) exhibits clear stylistic word features \citep{Alafnan2023}. We exploit a 65-term vocabulary previously linked to GenAI-related content \citep{Uribe2024,feyzollahi2025adoption,Filimonovic2025}. Table~\ref{tab:keywords} lists the keywords that are applied in stemmed form to detect GenAI use in scientific publications. 

Titles and abstracts are especially well-suited for detecting stylistic shifts \citep{PlavenSigray2017} and carry significant editorial weight \citep{Ketcham2010}, making them meaningful proxies for GenAI uptake in scientific writing. We readily acknowledge that GenAI tools are likely used throughout a publication, and thus suggest that our estimates represent a lower bound on GenAI adoption.

\begin{table}[htbp]
\centering
\caption{Lexical Patterns Used for Detecting GenAI-generated Text}
\label{tab:keywords}
\begin{tabularx}{\textwidth}{X}
\toprule
\textbf{Keywords and Stems} \\
\midrule
delv*, groundbreak*, intric*, meticul*, realm*, revolution*, showcas*, underscore*, unveil*, while, elevat*, valuabl*, crucial*, empower, unleash, unlock, lever*, embarked, relentless, endeavour, enlightening, insight*, esteemed, resonate*, enhanc*, expertise*, offering*, tapestry, foster*, systemic*, inherent, synerg*, explor*, pivotal*, adhere, amplif*, embark*, invaluabl*, enlighten*, conceptual*, emphasiz*, complexit*, recogniz*, adapt*, promot*, critique, comprehensive, implication*, complementar*, perspective*, holistic, discern, multifacet*, nuanc*, underpinning*, cultivat*, integral, profound*, facilitat*, encompass*, elucidat*, unravel*, paramount, characteriz*, significant* \\
\bottomrule
\end{tabularx}
\footnotetext{Note: Asterisks (*) indicate stemmed keywords (all variants sharing the same root are grouped together; e.g., `delv*' captures `delve', `delving', `delved').}
\end{table}

Recognizing that keyword occurrence may vary by scientific field, we apply a field-specific frequency-based filter that restricts the analysis to terms whose relative frequency increased by at least 300\%: i.e., at least quadrupled in use from 2021 to 2024 and measured separately within each scientific field. We employ a high threshold as more modest shifts could reflect unrelated field-specific usage rather than any GenAI tool assistance. This approach helps to better isolate those terms whose diffusion is plausibly associated with GenAI adoption. Appendix Figure~\ref{fig:GenAI_Word_Freq} presents the final selection of terms (denoted as keywords achieving at least a four-fold increase) by scientific field, and shows a clear upward trend in usage frequency from 2021 to 2024.

Scopus assigns scientific fields at the journal level and journals can be cross-listed in multiple fields. As a result, a single publication can belong to more than one scientific field. Because linguistic baselines differ across fields, however, we identify GenAI markers and thresholds within field-year strata and use the following "any-field" rule: A publication is classified as GenAI-assisted if it is flagged---i.e., at least one of the filtered keywords in stemmed form appears in its title or abstract---in any of its assigned fields. All remaining publications are treated as non-GenAI-assisted. This approach reduces false negatives and ensures a single, consistent label per publication. We nevertheless examine the empirical robustness of this classification approach by applying different frequency thresholds. More and less restrictive thresholds alter the final term composition used to identify GenAI-assisted publications, but do not affect our baseline findings by much. Section~\ref{sec:rob} provides discussion around and analysis of this threshold classification approach.

\subsection{Linguistic Similarity Measurement}

We use text embeddings to measure the linguistic similarity between publication titles and abstracts by authors with institution affiliations to non-English-speaking countries and by authors with U.S.\ institution affiliations only. Text embeddings are vector representations in a high-dimensional semantic space, where proximity reflects similarity in language use and in underlying content. Because embeddings capture both style and content, all similarity comparisons are conditioned on field and publication year, so the remaining variation reflects differences in expression within comparable scientific contexts. We use SciBERT rather than general BERT because it is pretrained from scratch on scientific text, employs a domain-specific vocabulary that represents technical terms more accurately, and better models the linguistic patterns typical of scientific writing \citep{beltagy2019scibert}.

For each publication, we construct an embedding by concatenating its title and abstract, tokenizing this text with the pretrained SciBERT (\texttt{scivocab, uncased}) model from Hugging Face \citep{beltagy2019scibert}, and processing it with SciBERT using its maximum sequence length of 512 subword tokens (shorter texts are padded and longer ones truncated). From the model's final hidden states, we derive a 768-dimensional mean-pooled text vector (averaging token embeddings across the sequence). This approach is widely used for capturing overall text semantics with BERT models \citep{reimers2019sentence,ortakci2024revolutionary}. Missing titles or abstracts are replaced with empty strings prior to concatenation to ensure consistent processing, but this affects only a negligible proportion of the sample. Publications missing field information are excluded from the analysis. Embeddings are computed one publication at a time without gradient updates or fine-tuning, and serve as the basis for our analysis. We quantify linguistic proximity by computing cosine similarities between each non-U.S.\ publication and every `pure' U.S.\ publication (i.e., only U.S.\ affiliations) from the same scientific field and publication year: e.g., a Swiss chemistry publication from 2022 is compared to pure U.S.\ chemistry publications from 2022. We use L2 normalized embeddings for this calculation and average the resulting pairwise similarities to obtain similarity scores relative to the U.S.\ benchmark corpus. Cosine similarity values range from 0 to 1, where higher values indicate greater linguistic expression and semantic content similarity. These similarity scores form the basis of our empirical analysis, in which we test whether GenAI-assisted publications from non-English-speaking countries show stronger convergence toward U.S.\ benchmarks post-2022 than non-GenAI-assisted publications from these same countries.

We demonstrate empirical robustness by preserving the same field and year match but varying the benchmark corpus of pure US publications as follows: (1)  only those in top journals (viz., top 10\% by field-specific journal impact factor); (2) only those non-GenAI-assisted; and (3)  only those from 2021 in the same scientific field (i.e., predating ChatGPT's release date). Section~\ref{sec:rob} provides the corresponding empirical results.

\subsection{Empirical Strategy}\label{sec:model}
We briefly present descriptive context of the final sample used in the analysis. Table~\ref{tab:descriptive} provides summary statistics for the main variables. Linguistic similarity to U.S.\ scientific writing averages 0.82 ($SD=0.03$), indicating high overall similarity with modest cross-country variation. The average journal impact factor is $6.2$, with substantial variation across outlets ($SD=5.5$). GenAI-assisted publications account for about six percent of all observations, with yearly distributions roughly balanced across 2021--2024. Most observations originate from Asia ($59$ percent) and Europe ($32$ percent); the modest Americas share (five percent) reflects our publication restriction of at least one author affiliation in a non-English-speaking country. Life Sciences ($32$ percent) and Physical Sciences ($38$ percent) represent roughly two-thirds of the sample, with Engineering and Technology ($23$ percent) and Social Sciences ($7$ percent) accounting for the remainder.

\begin{table}[htbp]
\centering
\caption{Selected Variable Descriptive Statistics}
\label{tab:descriptive}
\begin{tabular}{lrrrr}
\toprule
\textbf{Variable} & \textbf{Mean} & \textbf{SD} & \textbf{Min}
  & \textbf{Max} \\
\midrule
Linguistic similarity to U.S.\ benchmark & 0.82 & 0.03 & 0.29 & 0.90 \\
Journal impact factor                    & 6.22 & 5.5  & 0.1  & 435.4 \\
Number of authors                        & 7.20 & 8.87 & 1    & 102   \\
GenAI-assisted publication (share)       & 0.06 &      &      &       \\
\midrule
\multicolumn{5}{l}{\textit{Year indicators (mean = share of sample)}} \\
Year: 2021 & 0.23 & & & \\
Year: 2022 & 0.24 & & & \\
Year: 2023 & 0.26 & & & \\
Year: 2024 & 0.27 & & & \\
\multicolumn{5}{l}{\textit{Continent indicators}} \\
Africa   & 0.04 & & & \\
Americas & 0.05 & & & \\
Asia     & 0.59 & & & \\
Europe   & 0.32 & & & \\
Oceania  & 0.00 & & & \\
\multicolumn{5}{l}{\textit{Aggregated field indicators}} \\
Engineering \& Technology & 0.23 & & & \\
Life Sciences             & 0.32 & & & \\
Physical Sciences         & 0.38 & & & \\
Social Sciences           & 0.07 & & & \\
\bottomrule
\end{tabular}
\footnotetext{Note: This table reports descriptive statistics for the regression sample ($N = 14{,}315{,}948$ observations).
The unit of observation is the publication-country-field combination. For binary variables, only the mean is reported.}
\end{table}

\begin{figure}[!htbp]
\centering
 \includegraphics[width=0.7\textwidth]{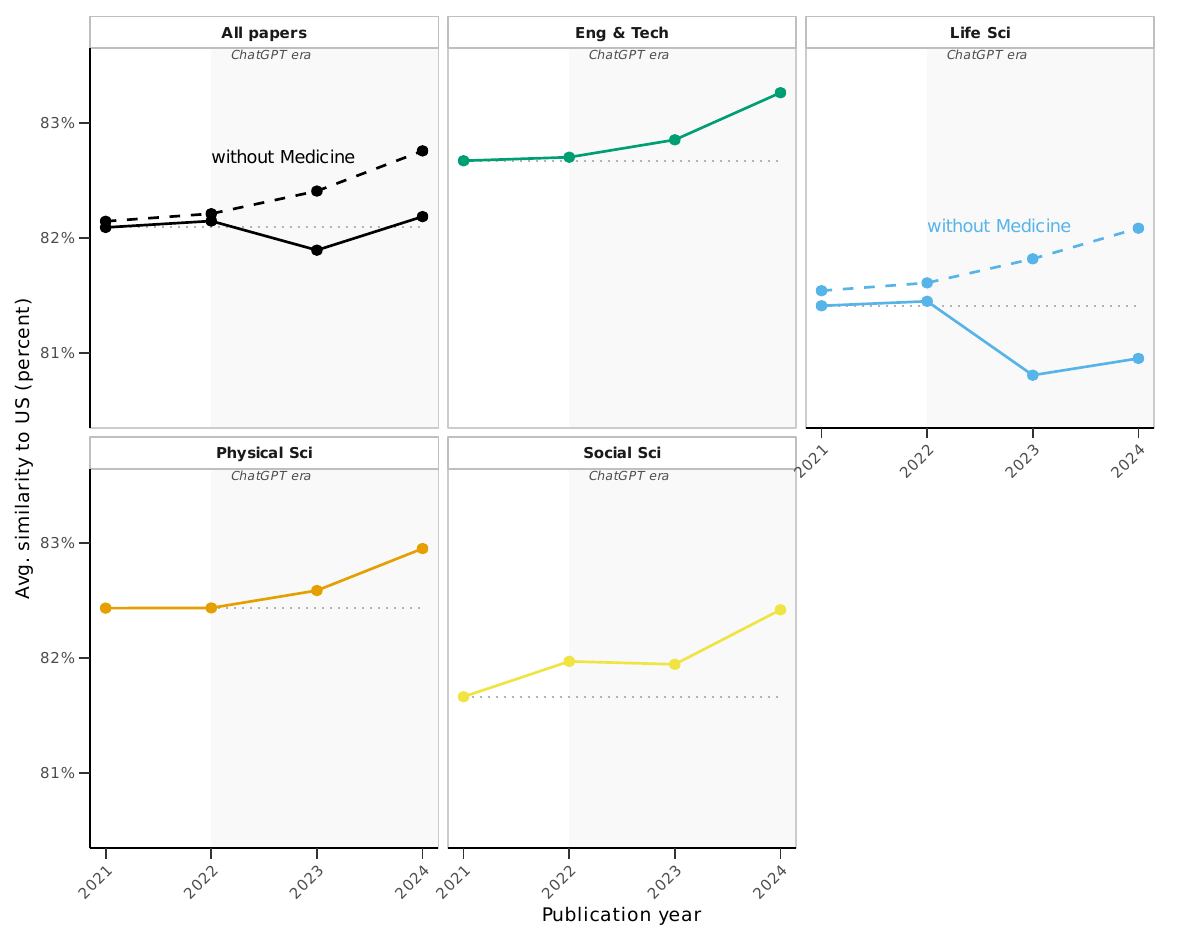}\par\vspace{0.5cm}
\caption{\textbf{Trends in Linguistic Similarity to U.S.\ Scientific Writing.}
This figure shows average linguistic similarity of publications from non-English-speaking countries to U.S.\ scientific writing, measured using SciBERT embeddings, and shown by publication year and aggregated scientific domain. Dashed lines denote values when publications in medicine are excluded.}
\label{fig:trends}
\end{figure}

Figure~\ref{fig:trends} shows linguistic similarity trends for publications from non-English-speaking countries relative to U.S.\ publications at aggregated scientific field levels. Similarity levels remain largely stable from 2021 to 2022, but rise significantly from 2023 onward as GenAI tools diffuse. Engineering and Technology experience the most pronounced linguistic similarity increase, while Life Sciences linguistic similarity was relatively stable through 2022 and then declined in 2023. This pattern is unlikely to reflect random variation, and instead suggests that the SciBERT-based measure captures both linguistic and topical dimensions. A plausible explanation is that COVID-19-related research temporarily raised linguistic similarity over 2021--2022, but as such research receded in 2023, Life Sciences publications returned to more typical topics and subsequently lowered linguistic similarity. Calculations that exclude medicine publications from Life Sciences support this explanation, revealing a notable upward trend.

The descriptive context is informative, but does not establish whether the observed linguistic convergence reflects GenAI use or broader research topic, field composition, or editorial practice shifts. We therefore turn to a formal econometric framework that compares GenAI-assisted and non-GenAI-assisted publications within the same journal and year, isolating stylistic shifts from concurrent compositional changes. Our empirical approach follows an event-study Difference-in-Differences (DiD) design with repeated cross-sections, comparing GenAI-assisted and non-GenAI-assisted publications to a U.S.\ benchmark corpus across calendar years and controlling for fixed effects to absorb time-invariant heterogeneity. We use 2022 as the reference year. We include pre-2023 years when widespread GenAI adoption was still limited to strengthen our design and create a placebo test: viz., if publications flagged by our linguistic markers over 2021-2022 do not differ from other publications in those years, it demonstrates that the observed shifts only emerge once GenAI is actually in use. The pre-GenAI period thus provides a critical baseline, helping disentangle changes driven by genuine GenAI adoption from those explained by the mere presence of or stylistic preferences for certain terms. The model specification is as follows:

\begin{equation}
\label{eq:main}
\begin{aligned}
y_{ifcjt} = \; &\delta_0 \cdot \text{GenAI}_{ifcjt}
  + \sum_{\tau \in \mathcal{T} \setminus \{2022\}}
    \beta_{\tau} \cdot \mathbf{1}\{t = \tau\}
    \times \text{GenAI}_{ifcjt} \\
  &+ \gamma_1 \cdot \text{Authors}_{ifcjt}
   + \gamma_2 \cdot \text{Eng}_{ifcjt} \\
  &+ \alpha_c + \alpha_f + \alpha_j + \alpha_t
   + \alpha_{j \times t} + \varepsilon_{ifcjt},
\end{aligned}
\end{equation}
\newline

\noindent where the dependent variable $y_{ifcjt}$ is our previously defined continuous similarity score for publication $i$ in field $f$, country $c$, journal $j$, and year $t$. $\text{GenAI}_{ifcjt}$ is an indicator equal to one if GenAI-related linguistic markers are detected and zero otherwise. $\tau$ represent year indicators, relative to 2022. The main coefficients of interest are $\beta_{\tau}$, which represent the interactions between the GenAI indicator and the year indicators and capture how the GenAI-assisted and non-GenAI-assisted publication similarity gap in year $\tau$ differs from the 2022 baseline. A positive (negative) value indicates that GenAI-assisted texts are converging (diverging) toward the U.S.\ benchmark, relative to 2022. We control at the publication level for the number of authors ($\text{Authors}_{ifcjt}$) and whether at least one coauthor is affiliated with an English-speaking country ($\text{Eng}_{ifcjt}$). We also include a rich set of fixed effects to control for unobserved heterogeneity: $\alpha_c$ are country fixed effects that absorb time-invariant country-level writing differences; $\alpha_f$ are Scopus field fixed effects (distinct from our aggregated groups) that absorb time-invariant field-specific writing styles; $\alpha_j$ are journal fixed effects that absorb outlet- and editorial-specific time-invariant factors; and $\alpha_t$ and $\alpha_{j \times t}$ are respective year and journal-by-year fixed effects that absorb common and journal-specific temporal shocks, such as shifts in the topics covered. Finally, $\varepsilon_{ifcjt}$ represents an idiosyncratic error term. Standard errors are robust and clustered at the journal level.

\section{GenAI Use and Convergence to U.S. English}

\subsection{Baseline Analyses}\label{sec:baselineresults}

We first estimate Equation~\eqref{eq:main} in a baseline analysis that includes all controls and fixed effects. Because U.S.\ scientific writing serves as an evolving empirical benchmark in the analysis, our estimates capture relative changes in linguistic similarity across countries rather than absolute changes in national writing.

\begin{figure}[htbp]
\centering
\includegraphics[width=0.7\linewidth]{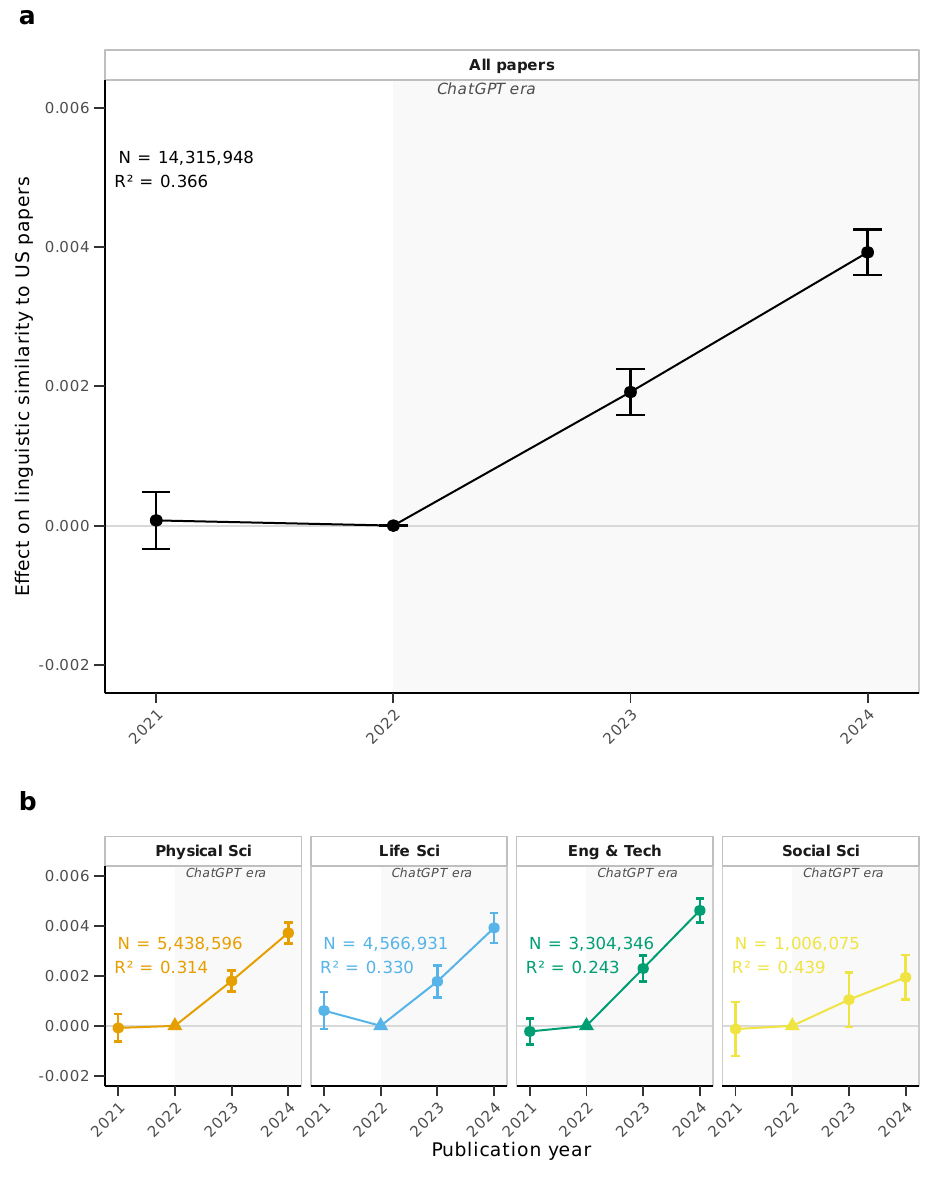}
\caption{\textbf{GenAI-Assisted Writing and Linguistic Convergence to U.S. Publications.}
This figure shows estimated effects based on the event-study regression (Equation~\ref{eq:main}), comparing GenAI-assisted and
non-GenAI-assisted publications from non-English-speaking countries. Coefficients represent year-specific differences in linguistic
similarity relative to the 2022 baseline. Panel (a) shows overall linguistic convergence in publications, while Panel (b) shows convergence by scientific field. All models include country, field, journal, year and journal-year fixed effects. Full regression results are provided in Appendix Table~\ref{tab:base}.}
\label{fig:eventstudy}
\end{figure}

Figure~\ref{fig:eventstudy} shows that the release of ChatGPT coincided with an increasing convergence in scientific writing styles--overall and by scientific field. Appendix Table~\ref{tab:base} provides full regression results. Panel (a) indicates that GenAI-assisted non-U.S.\ publications became more linguistically similar to the U.S.\ benchmark post-2022 than comparable non-GenAI-assisted publications. This overall effect is both highly statistically significant and increasing in magnitude: i.e., by 0.15\% in 2023 and by 0.4\% in 2024, relative to the 2022 baseline. The absence of any pre-trend suggests that this shift emerges post-launch of ChatGPT. Linguistic similarity scale magnitudes are small but as expected: viz., Figure~\ref{fig:trends} indicates that the average linguistic similarity already exceeds 0.8. Perhaps more informative than the absolute magnitudes is the clear directional shift that emerges with the post-launch of GenAI. Because the embedding measure can reflect both language and topical content shifts, we estimate models with country, field, journal, year, and journal-year fixed effects. Comparisons are thus made within the same year and journal----an empirical design that ensures the observed convergence among GenAI-assisted publications is unlikely to be driven by topical composition changes. The patterns remain highly consistent across robustness tests that vary both the identification of GenAI-assisted publications and the linguistic similarity specifications---including stricter keyword thresholds, alternative U.S.\ benchmark corpora, and the exclusion of GenAI-assisted U.S.\ texts (see Section~\ref{sec:rob} for full details and estimates).

Panel (b) indicates that this linguistic convergence is not confined to a single field  but widespread across the scientific landscape. In particular, consistent and relatively large upward trends are observed in Engineering and Technology (0.5\%), Physical Sciences (0.4\%) and Life Sciences (0.4\%), with a more modest increase in Social Sciences (0.2\%).

\subsection{Heterogeneity Analyses}\label{sec:heterogeneityresults}

\begin{figure}[htbp]
\centering
\includegraphics[width=0.9\linewidth]{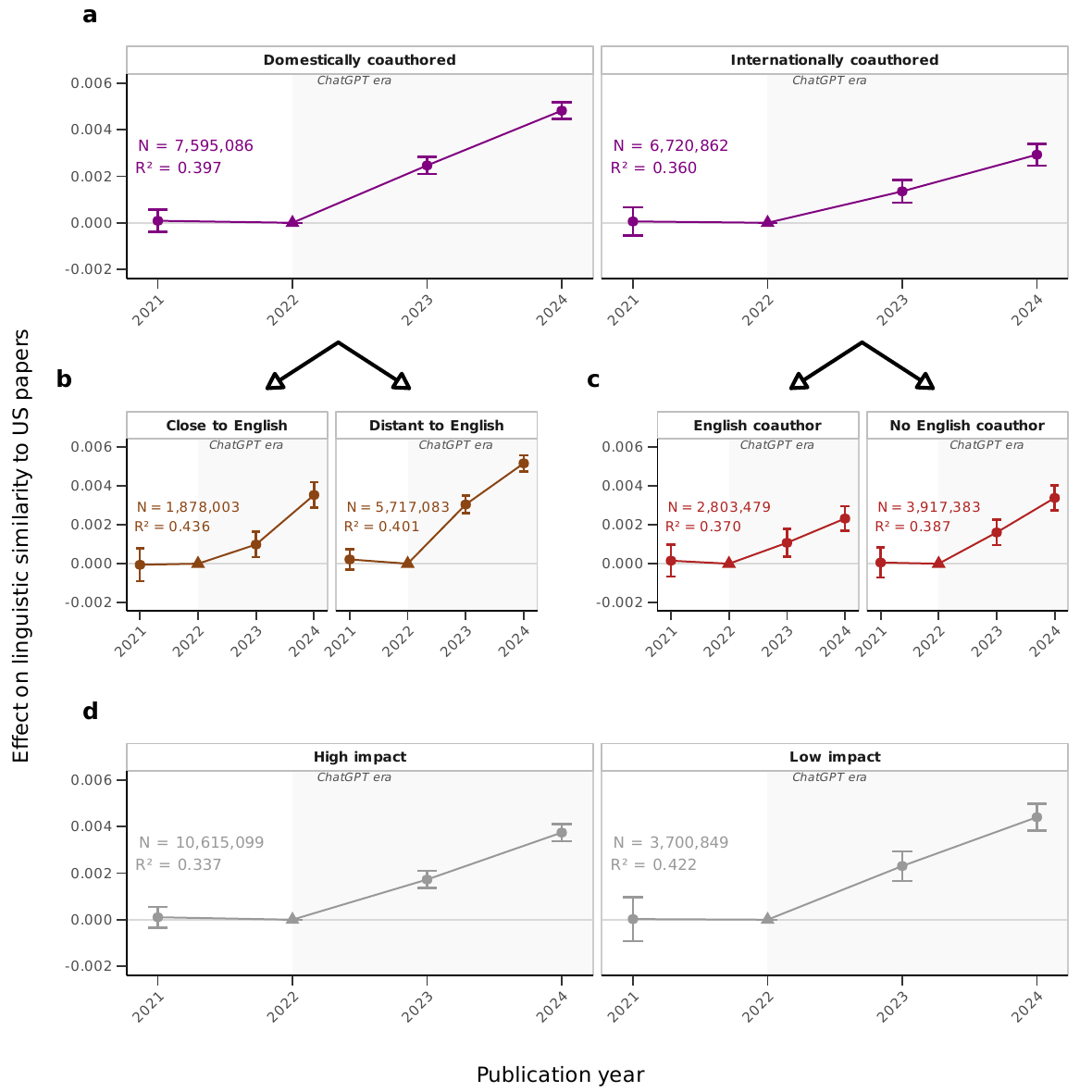}
\caption{\textbf{Linguistic Similarity Convergence: Subsample Estimations.}
This figure shows estimated effects based on the event-study regression (Equation~\ref{eq:main}), comparing GenAI-assisted to
non-GenAI-assisted publications. Panel~(a) contrasts domestically vs.\ internationally coauthored papers. Panel~(b) contrasts countries linguistically close vs.\ linguistically distant from English within domestically coauthored papers. Panel~(c) contrasts coauthor teams with vs.\ without an English-speaking coauthor. Panel~(d) contrasts publications in high-impact vs.\ low-impact
journals. All models include country, field, journal, year and journal-year fixed effects. Full regression results are provided in Appendix Table~\ref{tab:subsample}.}
\label{fig:heterogeneity}
\end{figure}

We next re-estimate Equation~\eqref{eq:main} in several heterogeneity analyses that consider other factors that might affect linguistic convergence. If GenAI acts as a converging force in scientific writing, its effects should be most pronounced for those researchers who have historically faced the greatest language-related publication barriers. Figure~\ref{fig:heterogeneity} provides results that are consistent with this prediction. Appendix Table~\ref{tab:subsample} provides full regression estimates.

Panel (a) compares domestic author team (all same-country authors) publications with international author team (at least one different-country author) publications. International author teams typically bring broader complementary capabilities (that plausibly include language support) and achieve higher impact \citep{Wuchty2007,adams2012rise,adams2013}, in comparison to domestic author teams. Consistent with this argument, we find that the convergence effect of GenAI is more pronounced for domestic author teams.

Panel (b) compares domestic author team publications from countries linguistically close to English to countries linguistically distant to English. Although English dominates international scientific communication, its distance to other languages varies considerably and likely shapes both the initial need for GenAI support and the potential for linguistic convergence. We capture this dimension using the Common Language Index (CLI) that aggregates three linguistic components \citep{Melitz2014}: (i) common native language (CNL) -- measuring the probability that two randomly chosen individuals from two countries share the same mother tongue; (ii) common official language (COL) -- capturing institutionalized support for communication via official status; and (iii) linguistic proximity (LP) -- quantifying similarities between basic vocabularies based on the Automated Similarity Judgment Program (ASJP) \citep{Wichmann2025asjp}. The resulting normalized score ranges from 0 to 1, where higher values indicate greater closeness to U.S.\ English. We split the sample at the median CLI score into proximate countries (above median) and distant countries (below median), and assess whether GenAI-driven linguistic convergence is stronger where researchers face larger inherent language barriers. Consistent with this argument, we find that the convergence effect is significantly larger and rises more sharply for linguistically-distant countries, in line with the interpretation that GenAI helps reduce language barriers for those facing greater baseline frictions.

Panel (c) contrasts international author team publications that include at least one English-speaking coauthor with those that include none. Because such teams likely face fewer language frictions \citep{Amano2023}, we expect smaller gains where English-speaking expertise is present. Consistent with this argument, we find stronger convergence for author teams without any English-speaking coauthor. 

Panel (d) considers this substitution mechanism across the scientific publication landscape: viz., low- versus high-impact journals. The results indicate that linguistic convergence is more prevalent in lower-impact journals---where authors may have less access to professional editing services---and less prevalent in top-tier journals---where manuscripts already undergo rigorous language polishing and perhaps face a ceiling effect. Previous evidence shows that peer review in high-impact journals is substantially more thorough---viz., reviews are longer and devote greater attention to substantive aspects, compared to lower-impact outlets \citep{Severin2023}. These results suggest that GenAI may disproportionately benefit researchers aiming to meet stylistic standards in outlets where language support is not institutionalized, and align with recent evidence from biomedicine indicating that less established researchers and those at lower-ranked institutions are the primary adopters of GenAI tools \citep{Liu2025}.

Across panels (a)--(c), GenAI appears to act as a partial substitute for human-based linguistic expertise---with the largest gains among domestic author teams from countries more linguistically distant from English. In panel (d), GenAI appears to act as a partial substitute especially in settings where language support is not institutionalized. Overall, the results consistently show that GenAI-assisted publications from non-English-speaking countries converge more rapidly toward U.S.\ scientific writing after 2022.

\section{Robustness} \label{sec:rob}

We conduct a series of empirical robustness tests to assess the stability of our baseline and subsample estimation results. These tests include alternative keyword thresholds, keyword threshold restrictions, and U.S. publication corpus definitions. 

\paragraph{Alternative Keyword Thresholds} The baseline estimation identifies GenAI-assisted publications using a field-specific frequency threshold of 300\% (i.e., occurrences increase by a factor of four between 2021 and 2024) for GenAI-related terms. We test whether this result is sensitive to this threshold level: viz., thresholds too low may produce too many false-positives; thresholds too high may produce too many false-negatives. We re-estimate the baseline specification using looser (i.e., 200\% or a three-fold occurrence increase) and stricter (i.e., 400\% or a five-fold occurrence increase) thresholds. 

Appendix Figure \ref{fig:GenAI_Word_Freq} displays GenAI-linked words whose frequency increased relative to their pre-ChatGPT baseline disaggregated by scientific field. The figure includes only those keywords from our initial list that meet our lowest threshold of a two-fold increase. The shading indicates the magnitude of the increase: i.e., more words are included under the looser (3x) threshold; fewer words are included under the stricter threshold (only those shaded 5× or higher). Identification is again conducted at the scientific field level, meaning words are selected according to their abnormal post-2022 frequency growth within each field. The looser criterion captures a broader set of linguistic markers, ensuring that our findings are not driven only by highly-distinctive cases. The stricter criterion focuses exclusively on the most pronounced GenAI-related terms. 

The left side of Appendix Figure \ref{fig:thresh_effect} illustrates how the share of GenAI-assisted publications varies across these thresholds: panel (a) uses the looser (200\%) threshold; panel (b) uses the baseline (300\%) threshold; and panel (c) uses the stricter (400\%) threshold. Although GenAI-assisted publications follow a similar pattern over time across the panels, the magnitude of GenAI adoption is roughly equivalent in panels (b) and (c) and markedly higher in panel (a). 
 
The right side of Appendix Figure \ref{fig:thresh_effect} presents estimation results of linguistic similarity convergence across these alternative keyword thresholds. For each threshold, the core result of greater convergence in linguistic similarity between non-U.S. GenAI-assisted publications and the U.S. corpus holds, relative to non-U.S. and non-GenAI-assisted publications. The variation in magnitudes across thresholds likely reflects differences in field composition and writing conventions among publications with stronger versus weaker GenAI signals. Importantly, these differences do not alter our main conclusion that convergence is robust to alternative keyword thresholds.

\paragraph{Alternative Keyword Threshold Restrictions} The baseline estimation classifies a publication as GenAI-assisted if at least one GenAI-linked keyword appears in its title or abstract. This definition maximizes sensitivity to early stylistic shifts, but may also capture publications where such language reflects author(s) unique writing style rather than genuine GenAI assistance. We examine this directly by implementing a stricter criterion that requires at least two distinct GenAI-related terms present. Publications flagged only by a single keyword are excluded from the sample. This robustness test ensures that our findings are not driven by marginal cases where GenAI-linked keywords appear without reflecting systematic adoption of GenAI writing support. Convergence that persists under this stricter definition provides additional support that the observed effects capture meaningful GenAI-related shifts in linguistic style rather than noise from spurious single-word matches. 

Appendix Figure \ref{fig:twokeyword} presents estimation results of linguistic similarity convergence, directly comparing the different identification approaches: i.e., the left panel provides the one-keyword results and the right panel provides the at least two-keyword results. Under the stricter identification definition, non-U.S. GenAI-assisted publications show a similar convergence---albeit lower in magnitude and delayed over time---to U.S.-publications, in comparison to the baseline identification definition. Notably, increasing convergence in linguistic similarity is not found in 2023 but is found in 2024, relative to the 2022 baseline. 

\paragraph{Alternative U.S. Publication Corpus Definitions} A critical assumption underlying our baseline estimation is that our set of U.S.-authored benchmark publications serves as an accurate and stable reference for "standard" scientific English. To ensure that our findings are not mechanically driven by how this benchmark is defined, we consider three alternatives. 

First, we re-compute similarity scores relative to a U.S. benchmark corpus from which publications flagged as GenAI-assisted have been excluded. This test is used to verify that convergence patterns are not an artifact of GenAI adoption in the benchmark itself, which could otherwise compress measured linguistic distances. Panel (a) in Appendix Figure \ref{fig:corpus_2021} presents estimation results that closely mirror those from our baseline estimation, indicating that the inclusion of GenAI-assisted papers in the benchmark corpus does not affect our findings.

Second, we restrict the benchmark to U.S. publications from 2021: that is, prior to the release of ChatGPT, and thereby providing a temporally-stable reference point that is unaffected by potential shifts in U.S. writing after 2022. Panel (b) in Appendix Figure \ref{fig:corpus_2021} presents the estimation results and shows a markedly similar pattern and pace of convergence over time compared with the baseline estimation.

Third, we construct a benchmark consisting of all U.S. publications appearing in the top 10\% of journals within each Scopus field, ranked by impact factor. This ensures that our results are not driven by average or lower-quality U.S. writing styles, and instead reflect convergence toward the linguistic register of leading publication outlets. Panel (c) in Appendix Figure \ref{fig:corpus_2021} presents the estimation results, and shows a similar pattern of convergence over time to the baseline estimation but with markedly larger magnitudes. This finding suggests that non-U.S. GenAI-assisted publications have converged more strongly in linguistic style to top-tier U.S. publications, indicating that the observed convergence is directed toward the style of leading outlets. However, the larger magnitudes may also partly reflect the greater linguistic homogeneity of top-journal publications, which provides a more concentrated benchmark against which convergence is easier to detect.  

These robustness tests help strengthen the credibility of our empirical strategy by demonstrating that the observed convergence of GenAI-assisted texts does not hinge on specific choices regarding the composition of our U.S. benchmark corpus.

\section{Discussion}

The convergence patterns documented above---strongest among domestic author teams from linguistically distant countries and in
lower-impact journals---suggest that GenAI acts as a partial linguistic equalizer in global science, narrowing the gap toward native English scientific writing norms, with disproportionately larger associations where linguistic and institutional resources are most
constrained. This interpretation is consistent with complementary evidence showing asymmetric productivity gains from LLM adoption
\citep{Kusumegi2025, Filimonovic2025} and readability improvements among non-native English-speaking authors \citep{Prakash2025}. And our analysis extends prior single-field or preprint-based analyses \citep{Lin2025,Liu2025} to published journal articles across four scientific domains.

The observed patterns raise important questions about potential downstream consequences for the global research ecosystem. Insights from international trade theory offer a useful framework to assess the impact of these shifts across multiple actors \citep{Melitz2003}. The most immediate beneficiaries are likely non-native English-speaking researchers who use GenAI tools to overcome language constraints, gaining greater visibility in international and higher-impact journals. Yet the influx of new contributors may intensify competition for native English-speaking incumbents, inducing sorting dynamics akin to firm-level responses to import competition \citep{Melitz2003,Amiti2013,Medina2024}: in particular, relatively `weaker' native English-speaking researchers may experience reduced success, whereas relatively `stronger' researchers may respond by further enhancing the quality and originality of their work. Broader participation may also expand the pool of ideas which future research can build on---analogous to improved access to imported `intermediate inputs' \citep{Amiti2007}.

Beyond this micro-level enrichment of the knowledge base, reducing language-related barriers may generate macro-level gains---much like trade openness is associated with aggregate welfare improvements through increased productivity, variety and quality \citep{Krugman1980,Feenstra1994,Melitz2003,Hummels2005}. A more level playing field may enable a wider range of high-quality ideas to compete and may gradually reduce the dominance of the Anglophone center \citep{amano2021tapping,nunez2021monolingual}. Over time, such diversification of contributors and ideas may help counteract the reported slowdown in disruptive discoveries and declining rates of new idea generation \citep{Bloom2020,Park2023}.

Realizing these gains will depend, however, on how journals, reviewers and policy-makers respond. For this potential to materialize, we suggest approaches around use and disclosure, equitable access, and editor and reviewer guidance: (i) To preserve trust without discouraging responsible GenAI use, journals could adopt transparent, low friction disclosure of language assistance and model use
\citep{crawford2023artificial,leung2023,thorp2023chatgpt}. For example, authors could clearly state what was used, for what tasks, and who verified the final content. (ii) To broaden access where marginal benefits are likely largest \citep{giglio2023}, policymakers and funding bodies could support equitable access and training for under-resourced institutions, including subsidized tools, shared infrastructure, and responsible use guidance. (iii) To better distinguish language clarity from scientific contribution, journals could update editorial and reviewer guidance during screening and review, with a goal of ensuring that declared
language support is treated as independent of scientific merit. In this context, resisting GenAI and leaning on LLM-based detectors
to prevent its use risks amplifying existing inequities. Current LLM-based detectors disproportionately flag non-native English
writing as `AI-generated' \citep{liang2023gpt}, and the rapid coevolution of writing styles in response to public awareness of LLM-characteristic vocabulary further undermines the reliability of lexical detection approaches \citep{Geng2024}---making such tools both unfair and ineffective. We therefore caution against their use in evaluative settings and emphasize policies that protect non-native English authors while promoting transparent and responsible language assistance.

Our study has limitations. First, our GenAI identification strategy relies on field-specific frequency shifts of lexical markers in titles and abstracts. Although we validate empirical robustness using stricter thresholds and alternative benchmarks (e.g., excluding GenAI-assisted U.S.\ publications, using a 2021 U.S.\ benchmark, and restricting to U.S.\ publications in high-impact journals), full-text stylometry and validated detection tools would further sharpen inference \citep{Hao2026}. Second, we classify authors affiliated with institutions in non-English-speaking countries as non-English, and those in the U.S.\ as English native. This distinction may not perfectly capture individual language proficiency, potentially introducing measurement noise. Third, similarity to a U.S.\ benchmark captures movement toward a dominant corpus of scientific English---not necessarily toward better science. Scholars have warned that such convergence may encourage stylistic homogenization and reduce rhetorical diversity  \citep{Flowerdew2022,House2003,Amano2023}, and experimental evidence indicates that AI writing suggestions can shift non-Western authors toward Western stylistic conventions \citep{Agarwal2025}. This possibility cannot be ruled out by our analysis and may warrant further investigation. Fourth, our design is observational. Linking author-level adoption to outcomes (e.g., acceptance rates, time to publication, citations, and reviewer text) and exploiting natural experiments (e.g., institutional rollouts and disclosure mandates) would strengthen causal claims. Finally, deeper understanding of field-specific norms, disclosure behavior, and  interactions between human editing and GenAI assistance will be essential to assess how language technologies reshape scientific practice.

\section{Conclusion}

This study set out to examine whether the diffusion of GenAI in scientific writing is associated with measurable linguistic convergence toward dominant benchmarks of scientific English. Specifically, we examine whether publications identified as GenAI-assisted exhibit stronger similarity to U.S.\ scientific writing than comparable non-GenAI-assisted publications, and whether such patterns vary systematically across linguistic and institutional contexts.

Using a large-scale bibliometric dataset of 5.65 million published papers and over 14 million unique publication-country-field observations (2021--2024), we document three main findings. First, GenAI adoption in scientific writing has been rapid but highly
uneven across countries, with higher uptake in linguistically distant contexts. Second, GenAI-assisted publications from non-English-speaking countries show statistically significant and increasing convergence toward U.S.\ scientific writing after 2022, relative to non-GenAI-assisted publications. Third, this convergence is strongest for those facing greater baseline language constraints---including all domestic author teams, all non-English-speaking author teams, and more linguistically-distant author teams---and for particular institutional constraints---such as in lower-impact journals. These patterns are robust to alternative keyword thresholds, stricter identification criteria, and different definitions of the U.S.\ benchmark corpus.

We make two methodological contributions and one conceptual contribution. Methodologically, we combine field-specific lexical markers with embedding-based similarity measures to provide a scalable framework for tracking stylistic shifts in published science. We also offer the first cross-field, cross-country evidence linking GenAI use to measurable changes in linguistic proximity to a dominant scientific benchmark. Conceptually, we position GenAI as a potential partial linguistic equalizer in global science, while highlighting that convergence toward a dominant register may also entail risks of stylistic homogenization \citep{Agarwal2025}.

The observational nature of our empirical design precludes causal inference, and our identification strategy captures stylized markers rather than direct usage logs. Nonetheless, the consistency of the convergence patterns across contexts and robustness checks suggests that GenAI is associated with systematic shifts in the linguistic presentation of scientific work. In sum, GenAI appears capable of reducing linguistic entry barriers in international publishing, potentially reshaping patterns of participation in ways that parallel the dynamics of trade liberalization and expanding the pool of competing ideas and contributors. Whether this potential translates into durable gains in scientific diversity and discovery will depend on complementary institutional investments, transparent disclosure norms, and thoughtful editorial and policy responses.

\backmatter

\bmhead{Acknowledgements}

We thank participants at the Institutions and Innovation Conference (IIC) for helpful
comments on earlier versions. Dragan Filimonovic and Christian Rutzer acknowledge general support from
the Zeaslin Bustany Foundation.

\section*{Declarations}

\paragraph*{Funding:}
No funding was received for conducting this study.

\paragraph*{Author contributions:}
Dragan Filimonovic contributed primarily to the conceptualization, writing, and empirical analysis, and assisted with data creation. Christian Rutzer contributed primarily to the conceptualization, writing, and data creation, and assisted with the empirical analysis. Jeffrey Macher contributed to the conceptualization and writing. Rolf Weder contributed to the conceptualization and writing.

\paragraph*{Competing interests:}
The authors have no relevant financial or non-financial interests to disclose.

\paragraph*{Data and materials availability:}
The data and code used in this study will be deposited on Zenodo and assigned a DOI upon publication. Raw Scopus data are subject to Elsevier licensing restrictions and cannot be shared publicly. Anonymized or aggregated data derived from Scopus, together with all analysis code, will be made available to enable replication.

\clearpage
\newpage
\begin{appendices}
\renewcommand{\theHtable}{A\arabic{table}}
\renewcommand{\theHfigure}{A\arabic{figure}}

\section{}\label{secA1}

\begin{figure}[!h]
  \centering
  \includegraphics[width=1\linewidth]{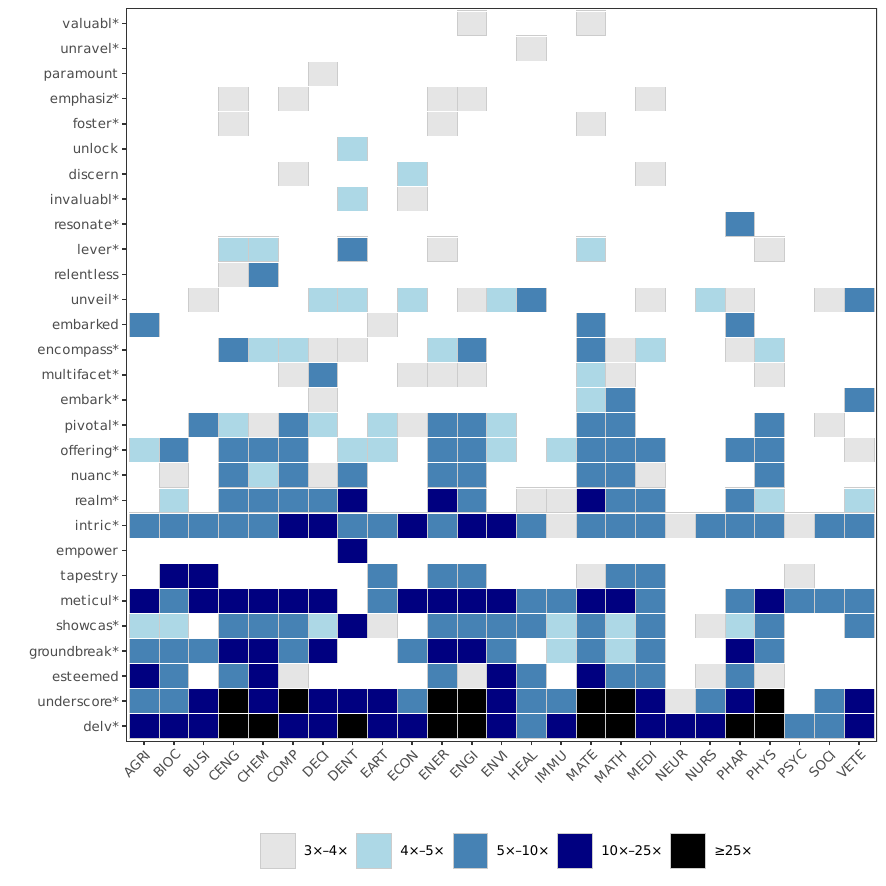}
  \caption{\textbf{Change in GenAI-Related Term Frequency Over Time and by Field.}
  This heatmap displays the change in the frequency of GenAI-related terms across scientific fields between 2021 and 2024. Each row corresponds to a linguistic pattern (GenAI-related term) and each column to a field. The shading indicates the fold-change in usage, with darker colors representing larger increases. Values are grouped into several change categories based on their magnitude ($3\text{--}4\times$, $4\text{--}5\times$, $5\text{--}10\times$, $10\text{--}25\times$, and $\geq25\times$). GenAI-related terms in Table~\ref{tab:keywords} that did not reach a three-fold increase in any field are excluded.}
  \label{fig:GenAI_Word_Freq}
\end{figure}

\begin{table}[]
\centering
\caption{GenAI-Assisted Writing and Linguistic Convergence to U.S. Publications}
    \label{tab:base}
    \footnotesize
\begin{tabular}{lccccc}
  \midrule \midrule
   Sample:                                              & (1)           & (2)           & (3)           & (4)           & (5)\\  
   \midrule
   \emph{Variables}\\
   Year 2021 × GenAI                                   & 0.000         & 0.000         & 0.001         & 0.000         & 0.000\\   
                                                       & (0.000)       & (0.000)       & (0.000)       & (0.000)       & (0.001)\\   
   Year 2023 × GenAI                                   & 0.002$^{***}$ & 0.002$^{***}$ & 0.002$^{***}$ & 0.002$^{***}$ & 0.001$^{*}$\\   
                                                       & (0.000)       & (0.000)       & (0.000)       & (0.000)       & (0.001)\\   
   Year 2024 × GenAI                                   & 0.004$^{***}$ & 0.005$^{***}$ & 0.004$^{***}$ & 0.004$^{***}$ & 0.002$^{***}$\\   
                                                       & (0.000)       & (0.000)       & (0.000)       & (0.000)       & (0.000)\\   
   GenAI paper                                         & 0.004$^{***}$ & 0.004$^{***}$ & 0.004$^{***}$ & 0.004$^{***}$ & 0.005$^{***}$\\   
                                                       & (0.000)       & (0.000)       & (0.000)       & (0.000)       & (0.000)\\   
   Number of authors                                   & 0.000         & 0.000$^{***}$ & 0.000$^{***}$ & 0.000$^{***}$ & 0.000\\   
                                                       & (0.000)       & (0.000)       & (0.000)       & (0.000)       & (0.000)\\   
   English-speaking coauthor & 0.003$^{***}$ & 0.003$^{***}$ & 0.004$^{***}$ & 0.003$^{***}$ & 0.003$^{***}$\\   
                                                       & (0.000)       & (0.000)       & (0.000)       & (0.000)       & (0.000)\\   
   \midrule
   \emph{Fixed-effects}\\
   Country                                             & Yes           & Yes           & Yes           & Yes           & Yes\\  
   Main field                                         & Yes           & Yes           & Yes           & Yes           & Yes\\  
   Journal                                             & Yes           & Yes           & Yes           & Yes           & Yes\\  
   Year                                                & Yes           & Yes           & Yes           & Yes           & Yes\\  
   Year-journal                                       & Yes           & Yes           & Yes           & Yes           & Yes\\  
   \midrule
   \emph{Fit statistics}\\
   Observations                                        & 14,315,948    & 3,304,346     & 4,566,931     & 5,438,596     & 1,006,075\\  
   R$^2$                                               & 0.366         & 0.243         & 0.330         & 0.314         & 0.439\\  
   Within R$^2$                                        & 0.007         & 0.012         & 0.008         & 0.011         & 0.005\\  
   \midrule \midrule
\end{tabular} 
\footnotetext{Note: The dependent variable is the linguistic similarity of a paper to U.S. scientific writing, measured using SciBERT embeddings. Column (1) reports results for the full sample of papers. Columns (2)--(5) report estimates for the respective scientific field groups: Engineering \& Technology, Life Sciences, Physical Sciences, and Social Sciences.\textit{GenAI} indicates papers that reference generative AI tools. Coefficients from the event-study specification capture differences between GenAI and non-GenAI papers by publication year relative to 2022 (reference year). All models include country, main field, journal, year, and year--journal fixed effects. Standard errors are clustered at the journal level and reported in parentheses. Significance levels: * $p<0.10$, ** $p<0.05$, *** $p<0.01$. }
\end{table}

\begin{landscape}

\begin{table}[]
    \centering
    \caption{Linguistic Similarity Convergence: Subsample Estimations.}
    \label{tab:subsample}
    \footnotesize
    \begin{tabular}{lcccccccc}
   \midrule \midrule
   Sample:                    & (1)           & (2)           & (3)           & (4)           & (5)           & (6)           & (7)           & (8)\\  
   \midrule
   \emph{Variables}\\
   2021 × GenAI              & 0.000         & 0.000         & 0.000         & 0.000         & 0.000         & 0.000         & 0.000         & 0.000\\   
                             & (0.000)       & (0.000)       & (0.000)       & (0.000)       & (0.000)       & (0.000)       & (0.000)       & (0.000)\\   
   2023 × GenAI              & 0.002$^{***}$ & 0.001$^{***}$ & 0.001$^{***}$ & 0.003$^{***}$ & 0.001$^{***}$ & 0.002$^{***}$ & 0.002$^{***}$ & 0.002$^{***}$\\   
                             & (0.000)       & (0.000)       & (0.000)       & (0.000)       & (0.000)       & (0.000)       & (0.000)       & (0.000)\\   
   2024 × GenAI              & 0.005$^{***}$ & 0.003$^{***}$ & 0.004$^{***}$ & 0.005$^{***}$ & 0.002$^{***}$ & 0.003$^{***}$ & 0.004$^{***}$ & 0.004$^{***}$\\   
                             & (0.000)       & (0.000)       & (0.000)       & (0.000)       & (0.000)       & (0.000)       & (0.000)       & (0.000)\\   
   GenAI paper               & 0.004$^{***}$ & 0.004$^{***}$ & 0.004$^{***}$ & 0.004$^{***}$ & 0.004$^{***}$ & 0.004$^{***}$ & 0.004$^{***}$ & 0.005$^{***}$\\   
                             & (0.000)       & (0.000)       & (0.000)       & (0.000)       & (0.000)       & (0.000)       & (0.000)       & (0.000)\\   
   Number of authors         & 0.000$^{***}$ & 0.000$^{***}$ & 0.000$^{***}$ & 0.000$^{***}$ & 0.000$^{**}$  & 0.000         & 0.000$^{**}$  & 0.000$^{***}$\\   
                             & (0.000)       & (0.000)       & (0.000)       & (0.000)       & (0.000)       & (0.000)       & (0.000)       & (0.000)\\   
   English-speaking coauthor &          & 0.002$^{***}$ &               &               &               &               & 0.003$^{***}$ & 0.003$^{***}$\\   
                             &    & (0.000)       &               &               &               &               & (0.000)       & (0.000)\\   
   \midrule
   \emph{Fixed-effects}\\
   Country                   & Yes           & Yes           & Yes           & Yes           & Yes           & Yes           & Yes           & Yes\\  
   Main field               & Yes           & Yes           & Yes           & Yes           & Yes           & Yes           & Yes           & Yes\\  
   Journal                   & Yes           & Yes           & Yes           & Yes           & Yes           & Yes           & Yes           & Yes\\  
   Year                      & Yes           & Yes           & Yes           & Yes           & Yes           & Yes           & Yes           & Yes\\  
   Year-journal             & Yes           & Yes           & Yes           & Yes           & Yes           & Yes           & Yes           & Yes\\  
   \midrule
   \emph{Fit statistics}\\
   Observations              & 7,595,086     & 6,720,862     & 1,878,003     & 5,717,083     & 2,803,479     & 3,917,383     & 10,615,099    & 3,700,849\\  
   R$^2$                     & 0.397         & 0.360         & 0.436         & 0.401         & 0.370         & 0.387         & 0.337         & 0.422\\  
   Within R$^2$              & 0.007         & 0.005         & 0.006         & 0.007         & 0.003         & 0.004         & 0.007         & 0.007\\  
   \midrule \midrule
\end{tabular}
\footnotetext{Note: The dependent variable is the linguistic similarity of a paper to U.S. scientific writing, measured using SciBERT embeddings. Columns (1)--(8) report estimates for the following groups: domestically coauthored papers (Column (1)), internationally coauthored papers (Column (2)), papers from countries linguistically close to English (Column (3)), papers from countries linguistically distant to English (Column (4)), papers with at least one English-speaking coauthor (Column (5)), papers without an English-speaking coauthor (Column (6)), papers published in high-impact journals (Column (7)), and papers published in low-impact journals (Column (8)). \textit{GenAI} indicates papers that reference generative AI tools. Coefficients from the event-study specification capture differences between GenAI and non-GenAI papers by publication year relative to 2022 (reference year). All models include country, main field, journal, year, and year--journal fixed effects. Standard errors are clustered at the journal level and reported in parentheses. The English-speaking coauthor indicator is omitted in specifications where it is collinear with the subsample definition. Significance levels: * $p<0.10$, ** $p<0.05$, *** $p<0.01$.}
\end{table}

\end{landscape}

\begin{figure}[!h]
  \centering
  \includegraphics[width=1\linewidth]{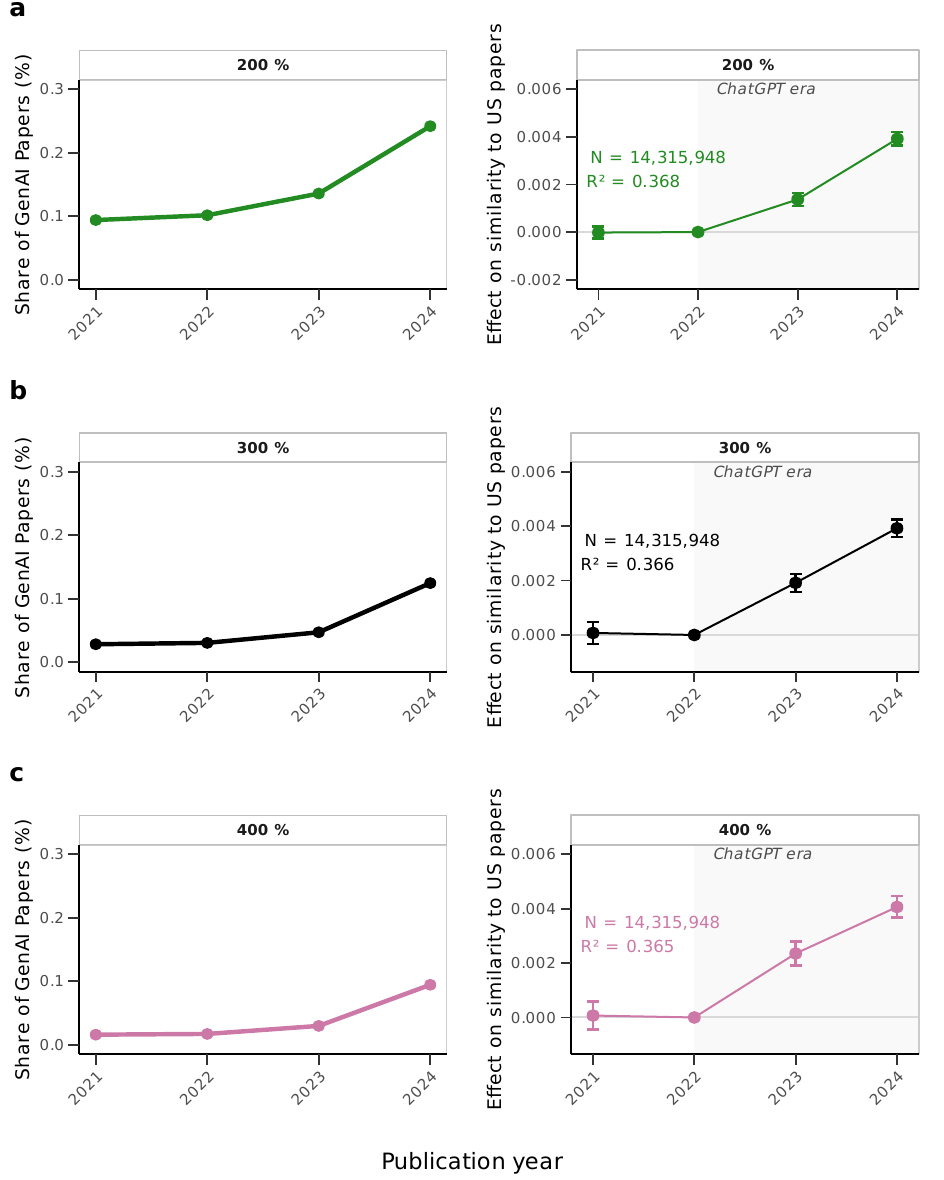}
  \caption{\textbf{Linguistic Similarity Convergence: Alternative Keyword Thresholds.}
  Panels (a)–(c) apply three alternative definitions for identifying GenAI-assisted publications based on changes in keyword frequency between 2021 and 2024: 200\% (a), 300\% (b; baseline), and 400\% (c) growth. In each panel, the left subfigure shows the evolution of the share of publications flagged as GenAI over time. The right subfigure reports coefficients from the event-study specification described in Materials and Methods. Estimated effects compare GenAI-assisted and non-GenAI-assisted publications originating from non English speaking countries. The dependent variable is the SciBERT-based linguistic similarity to U.S.\ scientific writing. Coefficients represent period-specific differences relative to the pre-GenAI baseline.}
  \label{fig:thresh_effect}
\end{figure}

\begin{figure}[!h]
  \centering
  \includegraphics[width=1\linewidth]{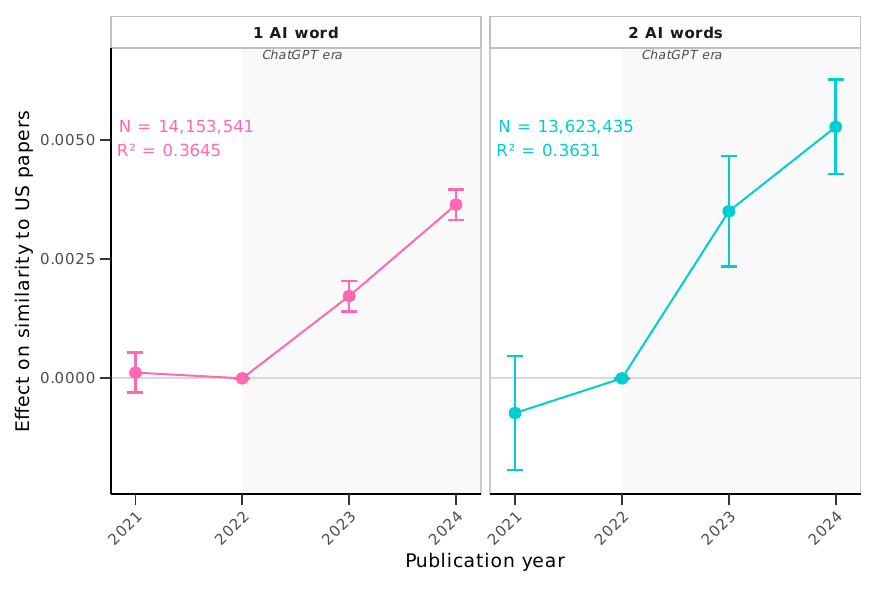}
  \caption{\textbf{Linguistic Similarity Convergence: Stricter Identification.}
  Both panels report coefficients from the specification described in Materials and Methods, i.e., our baseline model. The dependent variable is the SciBERT-based linguistic similarity to U.S.\ scientific writing, and estimates compare GenAI-assisted and non-GenAI-assisted publications from non English speaking countries relative to the pre-GenAI period. The left panel defines GenAI uptake based on the presence of exactly one GenAI-related term in either the title or the abstract. The right panel uses a stricter definition, flagging only papers that contain at least two such terms; papers with exactly one occurrence are excluded from both treatment and control groups.}
  \label{fig:twokeyword}
\end{figure}

\begin{figure}[!h]
  \centering
  \includegraphics[width=1\linewidth]{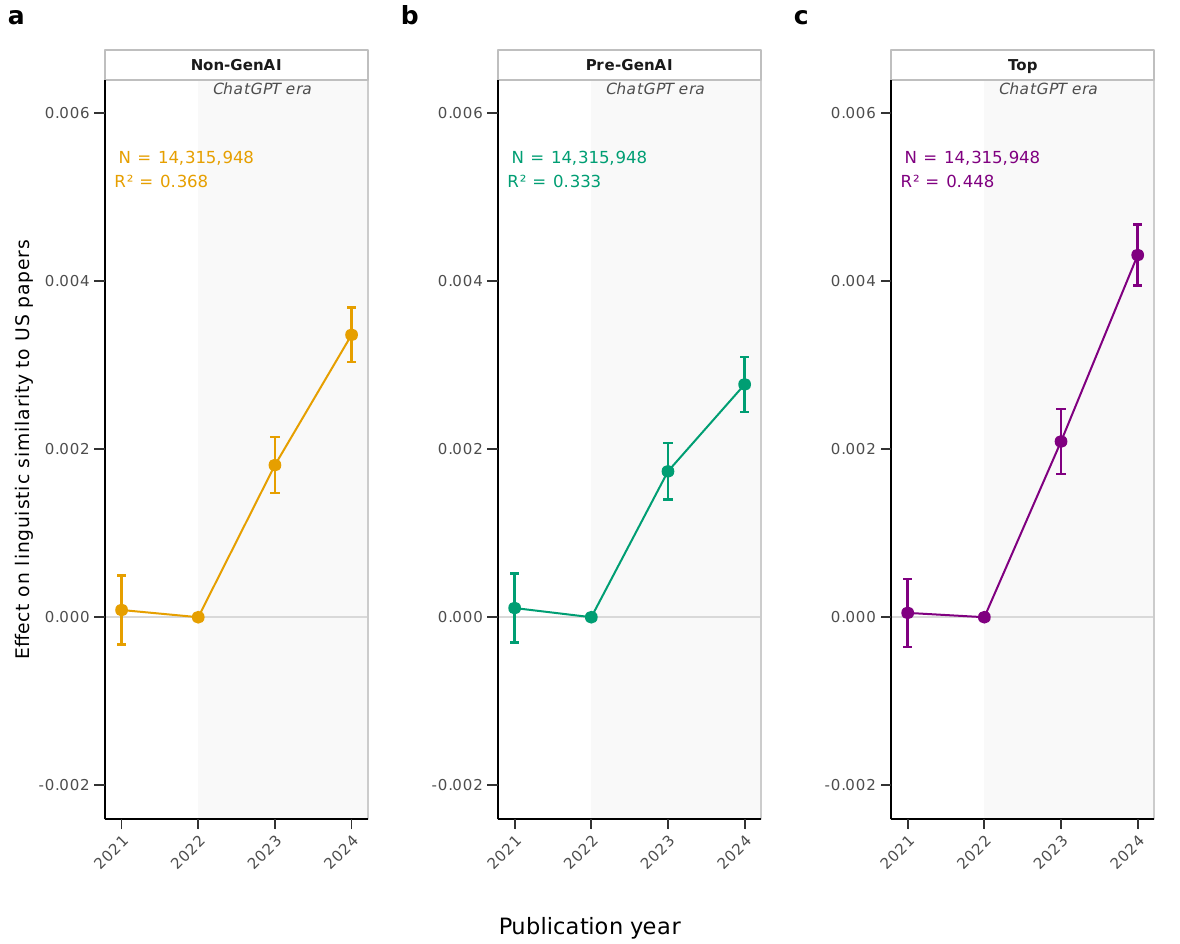}
  \caption{\textbf{Linguistic Similarity Convergence: Various U.S. Corpus Benchmarks.}
  Each panel reports estimates using the same specification described in Materials and Methods, i.e., our baseline model. The dependent variable is the SciBERT-based linguistic similarity to U.S.\ scientific writing, and coefficients compare GenAI-assisted and non-GenAI-assisted publications from non English speaking countries relative to the pre-GenAI period. The panels differ in how the U.S.\ benchmark corpus is defined: Panel (a) (\textit{Non-GenAI benchmark}) computes similarity scores relative to a U.S.\ corpus from which GenAI-assisted publications have been removed. Panel (b) (\textit{Pre-GenAI 2021 benchmark}) uses only U.S.\ publications from 2021, i.e., before the introduction of ChatGPT. Panel (c) (\textit{Top-US benchmark}) restricts the benchmark to U.S.\ publications appearing in journals that belong to the top 10 percent within each Scopus field, ranked by impact factor.}
  \label{fig:corpus_2021}
\end{figure}

\end{appendices}



\clearpage
\newpage

\begin{thebibliography}{66}
\providecommand{\natexlab}[1]{#1}
\providecommand{\url}[1]{{#1}}
\providecommand{\urlprefix}{URL }
\providecommand{\doi}[1]{\url{https://doi.org/#1}}
\providecommand{\eprint}[2][]{\url{#2}}
 \bibcommenthead

\bibitem[{Adams(2012)}]{adams2012rise}
Adams J (2012) The rise of research networks. Nature 490(7420):335--336

\bibitem[{Adams(2013)}]{adams2013}
Adams J (2013) The fourth age of research. Nature 497(7451):557--560

\bibitem[{Agarwal et~al.(2025)Agarwal, Naaman, and Vashistha}]{Agarwal2025}
Agarwal D, Naaman M, Vashistha A (2025) {AI} suggestions homogenize writing toward western styles and diminish cultural nuances. In: Proceedings of the 2025 {CHI} Conference on Human Factors in Computing Systems ({CHI} '25). ACM, pp 1--21

\bibitem[{Alafnan and Mohdzuki(2023)}]{Alafnan2023}
Alafnan MA, Mohdzuki SF (2023) {Do Artificial Intelligence Chatbots Have a Writing Style? An Investigation into the Stylistic Features of ChatGPT-4}. Journal of Artificial Intelligence and Technology 3(3):85--94

\bibitem[{Amano et~al.(2016)Amano, Gonz{\'a}lez-Varo, and Sutherland}]{Amano2016}
Amano T, Gonz{\'a}lez-Varo JP, Sutherland WJ (2016) Languages are still a major barrier to global science. PLoS biology 14(12):e2000933

\bibitem[{Amano et~al.(2021)Amano, Berdejo-Espinola, Christie, Willott, Akasaka, B{\'a}ldi, Berthinussen, Bertolino, Bladon, Chen et~al.}]{amano2021tapping}
Amano T, Berdejo-Espinola V, Christie AP, et~al (2021) Tapping into non-english-language science for the conservation of global biodiversity. PLoS Biology 19(10):e3001296

\bibitem[{Amano et~al.(2023)Amano, Ramírez-Castañeda, Berdejo-Espinola, Borokini, Chowdhury, Golivets, González-Trujillo, Montaño-Centellas, Paudel, White, and Veríssimo}]{Amano2023}
Amano T, Ramírez-Castañeda V, Berdejo-Espinola V, et~al (2023) The manifold costs of being a non-native english speaker in science. PLoS Biol 21(7)

\bibitem[{Amiti and Khandelwal(2013)}]{Amiti2013}
Amiti M, Khandelwal AK (2013) Import competition and quality upgrading. Review of Economics and Statistics 95(2):476--490

\bibitem[{Amiti and Konings(2007)}]{Amiti2007}
Amiti M, Konings J (2007) Trade liberalization, intermediate inputs, and productivity: Evidence from indonesia. American economic review 97(5):1611--1638

\bibitem[{Balan(2021)}]{Balan2021}
Balan S (2021) English as the language of research: But are we missing the mark? Exploratory research in clinical and social pharmacy 3:100043

\bibitem[{Barjak(2006)}]{Barjak2006}
Barjak F (2006) Research productivity in the internet era. Scientometrics 68(3):343--360

\bibitem[{Beltagy et~al.(2019)Beltagy, Lo, and Cohan}]{beltagy2019scibert}
Beltagy I, Lo K, Cohan A (2019) Scibert: A pretrained language model for scientific text. arXiv preprint arXiv:190310676 \unskip

\bibitem[{Bloom et~al.(2020)Bloom, Jones, Reenen, and Webb}]{Bloom2020}
Bloom N, Jones CI, Reenen JV, et~al (2020) Are ideas getting harder to find? American Economic Review 110(4):1104--1144

\bibitem[{Bubeck et~al.(2023)Bubeck, Chandrasekaran, Eldan, Gehrke, Horvitz, Kamar, Lee, Lee, Li, Lundberg, Nori, Palangi, Ribeiro, and Zhang}]{Bubeck2023}
Bubeck S, Chandrasekaran V, Eldan R, et~al (2023) Sparks of artificial general intelligence: Early experiments with gpt-4. \urlprefix\url{https://arxiv.org/abs/2303.12712}, {\href{https://arxiv.org/abs/2303.12712}{{arXiv:2303.12712}}}

\bibitem[{Castañeda(2020)}]{Castaneda2020}
Castañeda VR (2020) Disadvantages in preparing and publishing scientific papers caused by the dominance of the english language in science: The case of colombian researchers in biological sciences. PLoS ONE 15(9)

\bibitem[{Clavero(2010)}]{Clavero2010}
Clavero M (2010) "awkward wording. rephrase": linguistic injustice in ecological journals. Trends in Ecology \& Evolution 25(10):552--553

\bibitem[{Crawford et~al.(2023)Crawford, Cowling, Ashton-Hay, Kelder, and Middleton}]{crawford2023artificial}
Crawford J, Cowling M, Ashton-Hay S, et~al (2023) Artificial intelligence and authorship editor policy: Chatgpt, bard bing ai, and beyond. Journal of University Teaching and Learning Practice 20(5):1--11

\bibitem[{De~Haan et~al.(2025)De~Haan, Liu, Bollen, and Blanco}]{dehaan2025gpt}
De~Haan S, Liu Y, Bollen J, et~al (2025) Gpt editors, not authors: The stylistic footprint of llms in academic preprints. arXiv preprint arXiv:250517327 \unskip

\bibitem[{Devlin et~al.(2019)Devlin, Chang, Lee, and Toutanova}]{devlin2019bert}
Devlin J, Chang MW, Lee K, et~al (2019) Bert: Pre-training of deep bidirectional transformers for language understanding. Proceedings of the 2019 conference of the North American chapter of the association for computational linguistics: human language technologies, volume 1 (long and short papers) pp 4171--4186

\bibitem[{Elsevier(2023)}]{elsevier2023}
Elsevier (2023) The scopus content coverage guide: A complete overview of the content coverage in scopus and corresponding policies

\bibitem[{Feenstra(1994)}]{Feenstra1994}
Feenstra RC (1994) New product varieties and the measurement of international prices. American Economic Review pp 157--177

\bibitem[{Feyzollahi and Rafizadeh(2025)}]{feyzollahi2025adoption}
Feyzollahi M, Rafizadeh N (2025) The adoption of large language models in economics research. Economics Letters p 112265

\bibitem[{Filimonovic et~al.(2025)Filimonovic, Rutzer, and Wunsch}]{Filimonovic2025}
Filimonovic D, Rutzer C, Wunsch C (2025) Can genai improve academic performance? evidence from the social and behavioral sciences. arXiv preprint arXiv:251002408 \unskip

\bibitem[{Fitria(2021)}]{Fitria2021}
Fitria TN (2021) Grammarly as ai-powered english writing assistant: Students’ alternative for writing english. The Journal of English Language and Literature 5:65--78

\bibitem[{Flowerdew(2022)}]{Flowerdew2022}
Flowerdew J (2022) Models of english for research publication purposes. World Englishes 41(4):571--583

\bibitem[{da~Fonseca~Pachi et~al.(2012)da~Fonseca~Pachi, Yamamoto, da~Costa, and Lopez}]{Pachi2012}
da~Fonseca~Pachi CG, Yamamoto JF, da~Costa APA, et~al (2012) Relationship between connectivity and academic productivity. Scientometrics 93(2):265--278

\bibitem[{Geng and Trotta(2024)}]{Geng2024}
Geng M, Trotta R (2024) {Is ChatGPT Transforming Academics' Writing Style?} arXiv 2024. \urlprefix\url{http://arxiv.org/abs/2404.08627}, {\href{https://arxiv.org/abs/2404.08627}{{arXiv:2404.08627}}}

\bibitem[{Ghufron and Rosyida(2018)}]{Ghufron2018}
Ghufron MA, Rosyida F (2018) The role of grammarly in assessing english as a foreign language (efl) writing. Lingua Cultura 12(4):395--403

\bibitem[{Giglio and Costa(2023)}]{giglio2023}
Giglio AD, Costa MUPd (2023) The use of artificial intelligence to improve the scientific writing of non-native english speakers. Revista da Associacao Medica Brasileira 69(9):e20230560

\bibitem[{Gray(2024)}]{Gray2024}
Gray A (2024) Chatgpt "contamination": estimating the prevalence of llms in the scholarly literature. \urlprefix\url{https://arxiv.org/abs/2403.16887}, {\href{https://arxiv.org/abs/2403.16887}{{arXiv:2403.16887}}}

\bibitem[{Hanauer et~al.(2019)Hanauer, Sheridan, and Englander}]{Hanauer2019}
Hanauer DI, Sheridan CL, Englander K (2019) Linguistic injustice in the writing of research articles in english as a second language: Data from taiwanese and mexican researchers. Written Communication 36(1):136--154

\bibitem[{Hao et~al.(2026)Hao, Xu, Li, and Evans}]{Hao2026}
Hao Q, Xu F, Li Y, et~al (2026) Artificial intelligence tools expand scientists’ impact but contract science’s focus. Nature 649(8099):1237--1243

\bibitem[{House(2003)}]{House2003}
House J (2003) English as a lingua franca: A threat to multilingualism? Journal of Sociolinguistics 7(4):556--578

\bibitem[{Hummels and Klenow(2005)}]{Hummels2005}
Hummels D, Klenow PJ (2005) The variety and quality of a nation's exports. American Economic Review 95(3):704--723

\bibitem[{Ketcham et~al.(2010)Ketcham, Hardy, Rubin, and Siegal}]{Ketcham2010}
Ketcham CM, Hardy RW, Rubin B, et~al (2010) What editors want in an abstract. Laboratory Investigation 90:4--5

\bibitem[{Kobak et~al.(2025)Kobak, Gonz{\'a}lez-M{\'a}rquez, Horv{\'a}t, and Lause}]{kobak2025llm}
Kobak D, Gonz{\'a}lez-M{\'a}rquez R, Horv{\'a}t E{\'A}, et~al (2025) Delving into llm-assisted writing in biomedical publications through excess vocabulary. Science Advances 11(27)

\bibitem[{Krugman(1980)}]{Krugman1980}
Krugman P (1980) Scale economies, product differentiation, and the pattern of trade. American Economic Review 70(5):950--959

\bibitem[{Kusumegi et~al.(2025)Kusumegi, Yang, Ginsparg, {de Vaan}, Stuart, and Yin}]{Kusumegi2025}
Kusumegi K, Yang X, Ginsparg P, et~al (2025) Scientific production in the era of large language models. Science 390(6779)

\bibitem[{Leung et~al.(2023)Leung, de~Azevedo~Cardoso, Mavragani, and Eysenbach}]{leung2023}
Leung TI, de~Azevedo~Cardoso T, Mavragani A, et~al (2023) Best practices for using ai tools as an author, peer reviewer, or editor. Journal of Medical Internet Research 25:e51584

\bibitem[{Liang et~al.(2023)Liang, Yuksekgonul, Mao, Wu, and Zou}]{liang2023gpt}
Liang W, Yuksekgonul M, Mao Y, et~al (2023) Gpt detectors are biased against non-native english writers. Patterns 4(7)

\bibitem[{Liang et~al.(2026)Liang, Izzo, Zhang, Lepp, Cao, Zhao, Chen, Ye, Liu, Huang, McFarland, and Zou}]{Liang2024}
Liang W, Izzo Z, Zhang Y, et~al (2026) Monitoring ai-modified content at scale: A case study on the impact of chatgpt on ai conference peer reviews. \urlprefix\url{https://arxiv.org/abs/2403.07183}, {\href{https://arxiv.org/abs/2403.07183}{{arXiv:2403.07183}}}

\bibitem[{Lin et~al.(2025)Lin, Zhao, Tian, and Li}]{Lin2025}
Lin D, Zhao N, Tian D, et~al (2025) Chatgpt as linguistic equalizer? quantifying llm-driven lexical shifts in academic writing. arXiv preprint arXiv:250412317 \unskip

\bibitem[{Liu et~al.(2025)Liu, He, Zheng, Bu, and Ni}]{Liu2025}
Liu J, He Y, Zheng Z, et~al (2025) Ai-assisted writing is growing fastest among non-english-speaking and less established scientists. arXiv preprint arXiv:251115872 \unskip

\bibitem[{Medina(2024)}]{Medina2024}
Medina P (2024) Import competition, quality upgrading, and exporting: Evidence from the peruvian apparel industry. Review of Economics and Statistics 106(5):1285--1300

\bibitem[{Melitz and Toubal(2014)}]{Melitz2014}
Melitz J, Toubal F (2014) Native language, spoken language, translation and trade. Journal of International Economics 93(2):351--363

\bibitem[{Melitz(2003)}]{Melitz2003}
Melitz MJ (2003) The impact of trade on intra-industry reallocations and aggregate industry productivity. Econometrica 71(6):1695--1725

\bibitem[{Montgomery(2013)}]{Montgomery2013}
Montgomery SL (2013) Does science need a global language?: English and the future of research. University of Chicago Press

\bibitem[{Mundt and Groves(2016)}]{Mundt2016}
Mundt K, Groves M (2016) A double-edged sword: the merits and the policy implications of google translate in higher education. European Journal of Higher Education 6(4):387--401

\bibitem[{{Nature Human Behaviour}(2023)}]{nature2023scientific}
{Nature Human Behaviour} (2023) Scientific publishing has a language problem. Nature Human Behaviour 7:1019--1020

\bibitem[{Nuñez and Amano(2021)}]{nunez2021monolingual}
Nuñez MA, Amano T (2021) Monolingual searches can limit and bias results in global literature reviews. Nature Ecology \& Evolution 5(3):264--264

\bibitem[{Ortakci(2024)}]{ortakci2024revolutionary}
Ortakci Y (2024) Revolutionary text clustering: Investigating transfer learning capacity of sbert models through pooling techniques. Engineering Science and Technology, an International Journal 55:101730

\bibitem[{Park et~al.(2023)Park, Leahey, and Funk}]{Park2023}
Park M, Leahey E, Funk RJ (2023) {Papers and patents are becoming less disruptive over time}. Nature 613(7942):138--144

\bibitem[{Plav{\'e}n{-}Sigray et~al.(2017)Plav{\'e}n{-}Sigray, Matheson, Schiffler, and Thompson}]{PlavenSigray2017}
Plav{\'e}n{-}Sigray P, Matheson GJ, Schiffler BC, et~al (2017) The readability of scientific texts is decreasing over time. eLife 6:e27725

\bibitem[{Politzer-Ahles et~al.(2020)Politzer-Ahles, Girolamo, and Ghali}]{Politzer2020}
Politzer-Ahles S, Girolamo T, Ghali S (2020) Preliminary evidence of linguistic bias in academic reviewing. Journal of English for Academic Purposes 47:100895. \urlprefix\url{https://www.sciencedirect.com/science/article/pii/S1475158520301685}

\bibitem[{Prakash et~al.(2025)Prakash, Aggarwal, Varghese, and Varghese}]{Prakash2025}
Prakash A, Aggarwal S, Varghese JJ, et~al (2025) Writing without borders: Ai and cross-cultural convergence in academic writing quality. Humanities and Social Sciences Communications 12(1):1058

\bibitem[{Reimers and Gurevych(2019)}]{reimers2019sentence}
Reimers N, Gurevych I (2019) Sentence-bert: Sentence embeddings using siamese bert-networks. arXiv:190810084 \unskip

\bibitem[{Severin et~al.(2023)Severin, Strinzel, Egger, Barros, Sokolov, Mouatt, and Müller}]{Severin2023}
Severin A, Strinzel M, Egger M, et~al (2023) Relationship between journal impact factor and the thoroughness and helpfulness of peer reviews. PLOS Biology 21(8):e3002238

\bibitem[{Singh et~al.(2021)Singh, Singh, Karmakar, Leta, and Mayr}]{singh2021journal}
Singh VK, Singh P, Karmakar M, et~al (2021) The journal coverage of web of science, scopus and dimensions: A comparative analysis. Scientometrics 126:5113--5142

\bibitem[{Stephan et~al.(2016)Stephan, Franzoni, and Scellato}]{Stephan2016}
Stephan P, Franzoni C, Scellato G (2016) Global competition for scientific talent: Evidence from location decisions of phds and postdocs in 16 countries. Industrial and Corporate Change 25(3):457--485

\bibitem[{Tang et~al.(2025)Tang, Li, Hu, Zeng, and Du}]{Tang2025}
Tang C, Li SK, Hu S, et~al (2025) Gender disparities in the impact of generative artificial intelligence: Evidence from academia. PNAS Nexus 4(2):pgae591

\bibitem[{Thorp(2023)}]{thorp2023chatgpt}
Thorp HH (2023) Chatgpt is fun, but not an author. Science 379(6630):313--313

\bibitem[{Uribe and Maldupa(2024)}]{Uribe2024}
Uribe SE, Maldupa I (2024) {Estimating the use of ChatGPT in dental research publications}. Journal of Dentistry 149(July):105275

\bibitem[{Van~Noorden and Perkel(2023)}]{van2023}
Van~Noorden R, Perkel JM (2023) Ai and science: what 1,600 researchers think. Nature 621(7980):672--675

\bibitem[{Wichmann et~al.(2025)Wichmann, Holman, Brown, Dryer, and Ran}]{Wichmann2025asjp}
Wichmann S, Holman EW, Brown CH, et~al (eds)  (2025) {The ASJP Database (version 21)}

\bibitem[{Wuchty et~al.(2007)Wuchty, Jones, and Uzzi}]{Wuchty2007}
Wuchty S, Jones BF, Uzzi B (2007) The increasing dominance of teams in production of knowledge. Science 316(5827):1036--1039

\bibitem[{Xu and Reed(2021)}]{Xu2021}
Xu X, Reed M (2021) The impact of internet access on research output--a cross-country study. Information Economics and Policy 56:100914

\end{thebibliography}
\end{document}